# In situ GISAXS study of the growth of Pd on MgO(001)

F. Leroy[1,*], C. Revenant[1], G. Renaud[1] and R. Lazzari,[1,2]

[1]Département de Recherche Fondamentale sur la Matière Condensée/SP2M*/NRS, CEA Grenoble, 17 avenue des Martyrs, F-38054 Grenoble Cedex 9, France

[2]Groupe de Physique des Solides, CNRS – Universités Paris 6-7 UMR 7588,

2 Place Jussieu, F-75251 Paris Cedex 05, France

**Abstract**

The morphology of growing Pd nano-particles on MgO(001) surfaces have been investigated *in situ*, during growth, by grazing incidence small angle x-ray scattering, for different substrate temperatures. The 2D patterns obtained are quantitatively analyzed, and the average morphological parameters (shape, size) deduced. Above 650 K, the aggregates adopt their equilibrium shape of truncated octahedron, and the interfacial energy is deduced.

## 1. Introduction

---

* Corresponding author: Tel.: +33-4-38-78-57-61; FAX: +33-4-38-78-51-38; E-mail: fleroy@drfmc.ceng.cea.fr

Palladium nano-particles on MgO(001) surfaces are model systems for heterogeneous catalysis and metal/oxides interface [1,2]. Their catalytic properties depend on their atomic structure [3] as well as on their morphology. On this latter point the Grazing Incidence Small Angle X-Ray Scattering (GISAXS) technique has emerged in the last decade as an advantageous new tool to analyze the shape and distribution of aggregates on a substrate [3,4], mainly because it statistically probes the whole sample and is fast and non destructive. We have recently developed this technique *in situ*, in UHV, during molecular beam deposition [5]. It provides data extending far in reciprocal space with intensity variation over several orders of magnitude, thus showing many GISAXS features that could not be revealed in previous measurements. In this paper, we concentrate on typical 2D GISAXS pictures obtained during the growth of palladium on MgO(001) by molecular beam epitaxy at three temperatures (550, 650 and 740 K).

## 2. Experiments

In a GISAXS experiment, the incident X-ray beam impinges on a surface at a grazing angle $\alpha_i$ (chosen here equal to the critical angle for total external reflection). The scattered intensity is recorded on a plane as a function of the exit angle $\alpha_f$ with respect to the surface plane, and of the in-plane angle $\theta_f$. These angles allow defining the reciprocal space coordinates $q_z$ and $q_y$, respectively perpendicular and parallel to the surface [6].

The experiments were performed at the European Synchrotron Radiation Facility (ESRF) on the ID32 undulator beamline [7], delivering a monochromatic (0.1210 nm wavelength), doubly focused X-ray beam. A dedicated experimental setup was built in order to perform GISAXS *in situ*, in UHV, during MBE growth, without any window before the sample, thus avoiding background scattering [5]. The beam size at the sample location was 0.05 × 0.2 mm$^2$ (H × V). The UHV chamber, mounted on a diffractometer, had a base pressure of 10$^{-10}$ mbar. The scattering from the islands was collected on a one megapixel 16-bit X-ray CCD camera located ~1.2 m downstream from the sample. A tungsten beamstop was used to mask the transmitted and specularly reflected beams.

The (15 × 15 × 0.5 mm$^3$) MgO(001) substrates provided by Earth Chemical (Japan) were prepared following an homemade procedure resulting in MgO surfaces of high crystalline quality [8]. Pd (99.99% purity) was evaporated at a rate of 1 Å/min (calibrated *in situ* by a quartz microbalance) using an Omicron EFM4 e-beam bombardment deposition cell.

GISAXS measurements were performed during growth, for three substrate temperatures (500 K, 650 K and 740 K). The last deposits were analyzed by plane view TEM, with the carbon replica method.

## 3. Results

We concentrate below on the GISAXS analysis for four samples: 0.9 nm at 550 K; 0.1 and 1 nm at 650 K and 3.0 nm at 740 K. The main characteristics and results for these four samples are reported in Table I.

*3.1. GISAXS analysis*

The GISAXS analysis was performed with a computer program recently developed in our group [9], in the framework of the Distorted Wave Born Approximation (DWBA) [6] and the Local Monodisperse Approximation (LMA) :

$$I(\mathbf{q}) = S(\mathbf{q}) \times \overline{F(\mathbf{q})^2} \qquad (1)$$

where $F(\mathbf{q})$ is the form factor of an island and $S(\mathbf{q})$ is the interference function, Fourier Transform of the (in-plane) island-island pair correlation function.

The GISAXS data were first fitted using different model interference functions [9], but none of them was found appropriate. We thus resorted to digitalize large-scale TEM plane views to define an *ad hoc* interference function, with two parameters: *D,* the average center to center distance between neighboring islands, and $\omega$, a disorder parameter. This function, determined on the final deposits, was found appropriate for all deposits.

Qualitatively, the experimental GISAXS patterns present two large scattering lobes visible along the parallel direction (separated by the specular Crystal Truncation Rod, partly hidden by the beam stop). The extent of the intensity parallel (resp. perpendicular) to the surface is inversely proportional to the average lateral size (resp. height) of the island. The separation between the two main lobes is inversely proportional to the average separation between neighboring islands. In order to analyze a 2D GISAXS picture, two cuts of the intensity in the ($\theta_f$, $\alpha_f$) plane are simultaneously fitted. One cut is parallel to the surface at the $\alpha_f$ (or $q_z$) position of maximum intensity and one is perpendicular to it at the $\theta_f$ (or $q_y$) position of the interference maximum. Then, a 2D GISAXS pattern is simulated with the parameters obtained from the previous fits.

3.2. *Pd/MgO(001) at 550 K*.

For the final (0.9 nm thick) deposit at 550 K, TEM shows that most islands adopt the cube/cube epitaxial relationship but do not have a precise geometrical shape. Owing to the TEM plane views and the GISAXS perpendicular asymptotic behavior, the mean island has a sphere-based shape [12]. The morphological parameters obtained by GISAXS are in excellent agreement with those found with the TEM technique (cf. Table 1), except for the inter-island distance.

*3.3. Pd/MgO(001) at 650 K.*

For the final (3.0 nm thick) deposit at 650 K, the TEM study indicates a very good (001) island epitaxy with well-defined shapes of truncated pyramids for the smallest and octahedron for the largest.

For the 0.1 nm thick Pd/MgO(001) deposit at 650 K, the GISAXS data and analysis are presented in Fig. 1. Since no TEM data was recorded for this deposit, the mean island shape used to fit the GISAXS, a truncated pyramid with a square base, was deduced from other microscopy works for similar temperature and island size conditions [2]. A very good agreement is obtained between simulated and experimental data. Note the very small size distribution deduced from the fits (cf. Table I) for this small amount of deposited material.

For the 1 nm thick Pd/MgO(001) deposit at 650 K, the 2D GISAXS was recorded as a function of azimutal angle, revealing strong rods of scattering along the {111} directions, together with a second order of scattering perpendicular to the surface. These data clearly pointed to a truncated octahedron-like average shape, which indeed reproduces very well the experimental GISAXS data.

*3.4. Pd/MgO(001) at 740 K.*

For the final (3.0 nm thick) deposit at 740 K, TEM shows a largely predominant (001) island epitaxy and the islands are (111), (001) and (110) faceted. The mean island shape was also assumed to be a truncated octahedron with a square base. Fig. 2 shows that the experimental 2D GISAXS picture is again very well reproduced. At this temperature, condensation is clearly incomplete as the island volume is slightly larger than at 650K despite a factor 3 in the nominal thickness. This observation is in agreement with previous work [12]

## 4. Discussion

Let us compare the dimensional parameters deduced from GISAXS and from TEM for the 0.9 nm thick Pd at 550 K and the 3.0 nm thick Pd at 740 K. For the latter case, the agreement is excellent for all parameters, unambiguously demonstrating the adequacy of the GISAXS measurements and quantitative analysis. For the former case, the agreement is good as concerns the size distribution parameter $\sigma_R$. However, the inter-island distance and radius deduced from GISAXS are approximately 20% smaller than those deduced from TEM. We believe that this difference is related to non compact particle shapes for this low temperature deposit. Representing these diverse shapes by a simple spherical shape is a crude approximation. In addition, TEM probes only a very small portion of the surface, as opposed to GISAXS, and the differences might arise from inhomogeneity of the island distribution on the surface.

Particularly interesting is the 1nm-thick deposit at 650 K, for which it can be shown that the islands have reached their equilibrium shape [6,11], a truncated octahedron with a square base. This allows us to use the Wulff-Kaishew construction to deduce the interfacial energy $\beta$, related to the aspect ratio as:

$$\beta = 2\sigma_{001}\left[1 - \frac{H}{2R} \times \frac{\sigma_{111}}{\sigma_{001}} \times \frac{1}{\sin(\theta)}\right], \quad (2)$$

where H (resp. R) are mean height (resp. edge) values, $\sigma_{001}$ = 1.64 J/m$^2$ ($\sigma_{111}$ = √3/2 $\sigma_{001}$, expected from the broken bond (pairwise interaction) model [13]) is the surface specific energy of the (001) [respectively (111)] facet and $\theta$ = 54.7° is the angle between (001) and (111) facets [14,15]. This yields $\beta \approx$ 1.1 J/m$^2$, which compares well with the value of 0.947 J/m$^2$ deduced from contact angle measurements of a liquid Pd droplet on MgO(001) and to a recent experimental value of 0.91 J/m$^2$ [16,17]. Thus under adequate conditions, GISAXS may give access, non-destructively, to the equilibrium shape of the islands.

## 5. Conclusions

We have shown on four Pd/MgO(001) deposits that the in situ 2D-GISAXS technique applied *in situ*, in UHV during growth, can yield quantitative measurements allowing to determine the average shape of the aggregates, their average dimensional parameters as well as their distributions. For two of them, these parameters have been compared to those obtained from TEM and a good agreement is found. We can conclude that GISAXS data analysis is validated which gives the opportunity to investigate nucleation and growth processes in details. These GISAXS data have also been used to deduce the interfacial energy between the metal and the oxide.

**Figure captions**

Fig. 1. 2D GISAXS intensity for a 0.1 nm thick Pd/MgO(001) deposit at 650 K: (a) experimental pattern; (b) 2D pattern simulated with the parameters obtained from the parallel and perpendicular fits and reported in Table I. The intensity is represented with a logarithmic scale, the $\theta_f$ and $\alpha_f$ axes range from 0 to 3°.

Fig. 2. 2D GISAXS experimental intensity for a 3.0 nm thick Pd/MgO(001) deposit at 740 K. (a) Experimental pattern. (b) Continuous line: cut of the experimental pattern parallel to the surface, along the horizontal black line in Fig. 2a; filled squares: best fit of the experimental cut. (c) Same as (b) but cut perpendicular to the surface along the vertical black line in Fig. 2a. (d) 2D pattern simulated with the parameters obtained from the parallel and perpendicular

fits. The intensity is represented with a logarithmic scale, the $\theta_f$ (resp. $\alpha_f$) axis ranges from 0 to 2.5° (resp. 3.0°).

**Table**

| $T$ (K) | $\varepsilon$ (nm) | $D$ (nm) | $D_{TEM}$ (nm) | $<R>$ (nm) | $R_{TEM}$ (nm) | $\sigma_R$ | $\sigma_{R\ TEM}$ | $<H>$ (nm) | $\sigma_H$ |
|---|---|---|---|---|---|---|---|---|---|
| 550 | 0.9 | 6.17±0.06 | 7.9±1 | 1.66±0.05 | 2.0±0.4 | 1.3±0.05 | 1.3±0.1 | 2.06±0.03 | 1.05±0.05 |
| 650 | 0.1 | 16.22±0.2 | - | 1.43±0.05 | - | 1.15±0.05 | - | 1.78±0.02 | 1.1±0.05 |
| 650 | 1.0 | 16.02±0.2 | - | 7.3±0.2 | - | 1.24±0.05 | - | 5.71±0.1 | - |
| 740 | 3.0 | 18.0±0.2 | 18.0±3 | 7.5±0.2 | 7.5±1 | 1.25±0.05 | 1.3±0.1 | 6.41±0.1 | 1.1±0.05 |

Table 1. GISAXS results from best fits for Pd deposited on MgO(001) at different temperatures $T$ and with various thicknesses $\varepsilon$. $D$ and $D_{TEM}$ are the inter-island distances obtained by GISAXS and TEM, respectively. $<R>$ obtained by GISAXS is the average radius for a sphere-based island and the maximum half lateral size for an octahedron with a square base; $<H>$ is the average height over the height distribution. The TEM radius $R_{TEM}$ is that of a disk of equivalent surface. Lognormal distributions of sizes are used with width parameter $\sigma_R$ for the lateral size distribution and $\sigma_H$ for the vertical one.

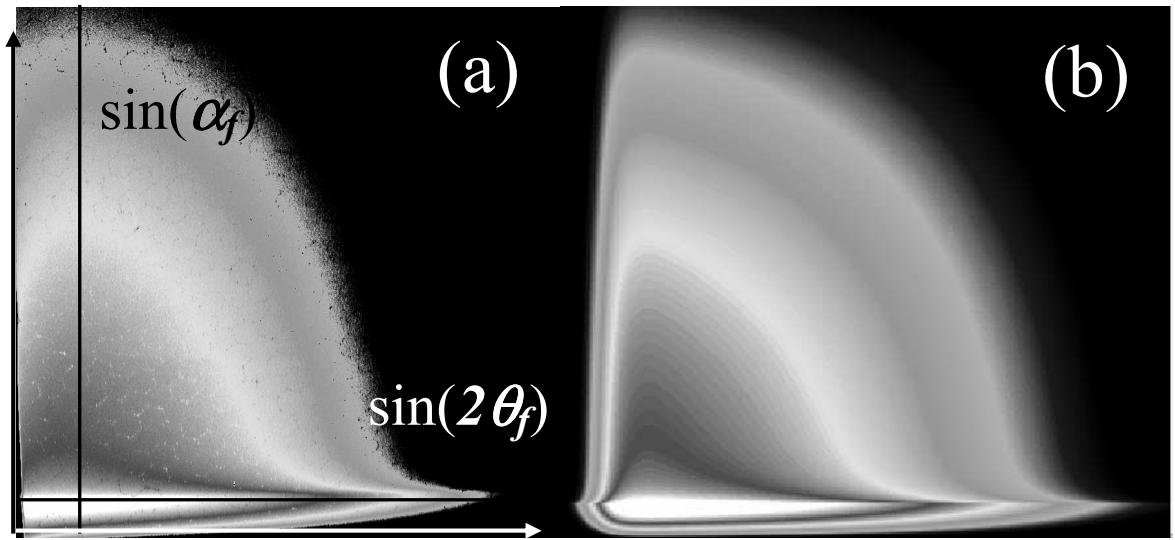

Fig. 1.

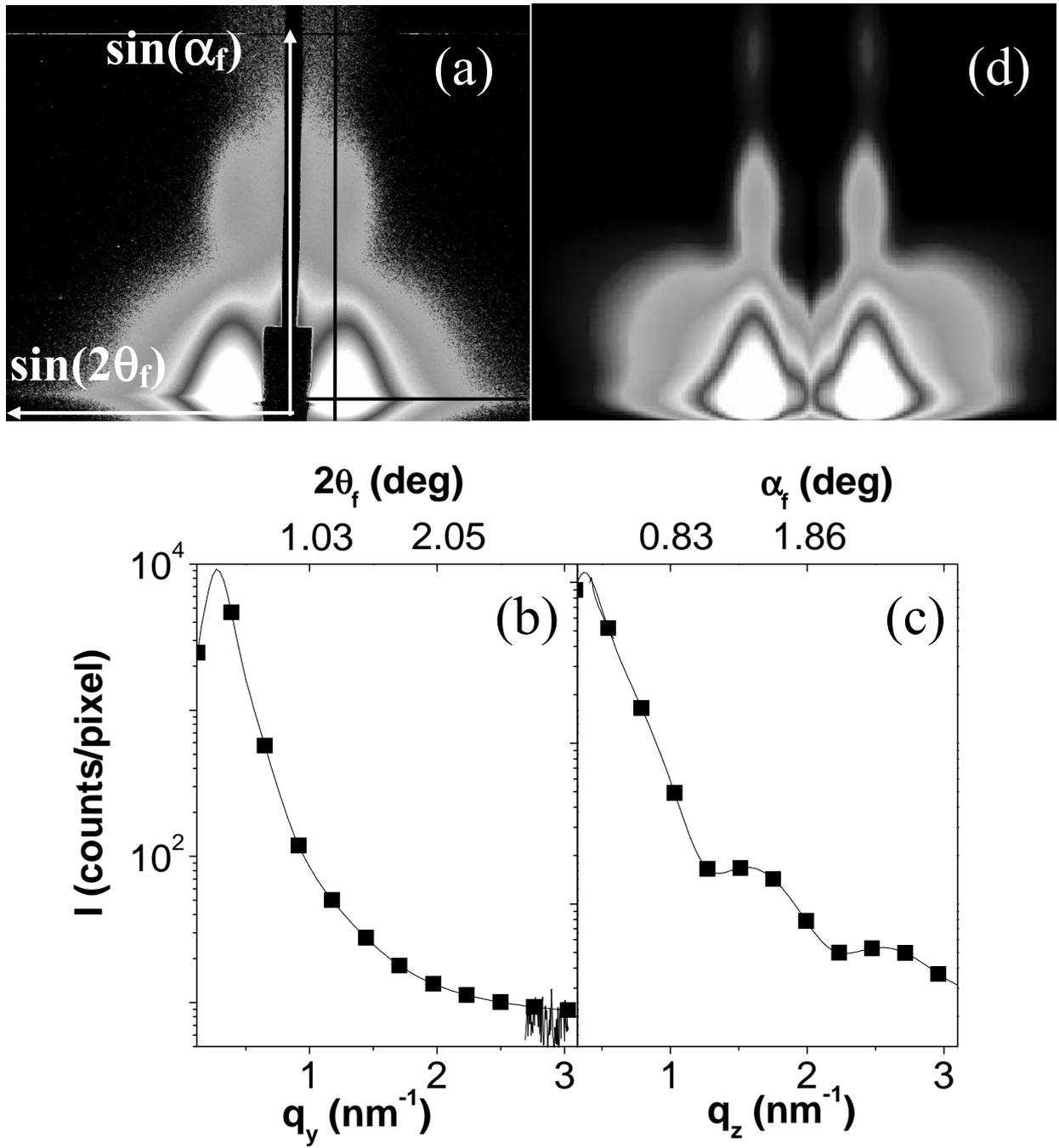

Fig. 2.